# Femtomolar Biodetection by a Compact Core-Shell 3D Chiral Metamaterial

Mariachiara Manoccio[1,2], Marco Esposito[1,*], Elisabetta Primiceri[1,*], Angelo Leo[1], Massimo Cuscunà[1], Vittorianna Tasco[1,*], Dmitry Zuev[3], Yali Sun[3], Giuseppe Maruccio[1,2], Alessandro Romano[4], Angelo Quattrini[4], Giuseppe Gigli[1,2] and Adriana Passaseo[1]

1 CNR NANOTEC Institute of Nanotechnology, Via Monteroni, 73100 Lecce, Italy
2 Department of Mathematics and Physics Ennio De Giorgi, University of Salento, Via Arnesano, 73100 Lecce, Italy
3 ITMO University, Department of Physics and Engineering, 49 Kronverkskiy av., St. Petersburg 197101, Russia
4 Division of Neuroscience, Institute of Experimental Neurology, San Raffaele Scientific Institute, Milan, Italy.
* Correspondence: vittorianna.tasco@nanotec.cnr.it, marco.esposito@nanotec.cnr.it, elisabetta.primiceri@nanotec.cnr.it

**Abstract.** Highly sensitive and selective label free devices for real-time identification of specific biomarkers are expected to significantly impact the biosensing field. The ability of plasmonic systems to confine the light in nanometer volume and to manipulate it by tuning the size, shape and material features of the nanostructures, makes these systems promising candidates for biomedical devices. In this work we demonstrate the engineered sensing capabilities of a compact array of 3D metal dielectric core-shell chiral metamaterial. The intrinsic chirality of the nano-helices makes the system circular polarization dependent and unaffected by the background interferences, allowing to work even in complex environment. The core-shell architecture enhances the sensing properties of the chiral metamaterial on both in the far and near field, also offering a large surface to molecular immobilization. With our system we recorded sensitivity of about 800nm/RIU and FOM= 1276 RIU$^{-1}$. The sensing abilities of the system is demonstrated with the detection of the TAR DNA-binding protein 43 (TDP-43), a critical biomarker for the screening of neurodegenerative diseases. In particular, the sensor was tested in different environments, such as human serum, with concentrations ranging from 1pM down to 10fM, opening new perspectives for novel diagnostic tools.

In biomedical research and clinical practice there is a huge demand of high performance sensors for advanced diagnostics and real-time monitoring of disease evolution[1,2]. Continuous technological improvements are required to increase sensitivity, specificity, and accuracy towards the challenging detection of small-size molecules or ultra-low concentration target analytes. In the last years, optical devices exploiting plasmonic properties emerged as good candidates to be implemented as low-cost, miniaturized and multiplexing biosensors[3–8].

Plasmonic biosensors are mostly based on the surface plasmon resonance (SPR), along metal/dielectric interfaces, and on the localized surface plasmon resonance (LSPR) occurring in nanostructures. Both detection mechanisms are strongly sensitive to the refractive index changes of the surrounding medium, within their respective plasmon decay lengths. Generally, SPR sensors exhibit higher sensitivity than LSPR, but require more complex excitation optics (such as prism or grating coupling) and extended smooth surfaces.

In order to improve the sensor performances one strategy consists in the material and shape engineering of plasmonic nanostructures[9–11]. In particular, the addition of anisotropy to plasmonic nanoobjects could enable novel degrees of freedom and polarization dependence in the optical response and, consequently, new biosensing concepts. One example are chiral plasmonic nanostructures[12,13], characterized by the absence of the mirror symmetry, which exhibit handedness and behave differently when interacting with circularly polarized light (CPL). In this respect, circular dichroism (CD) defined as the differential absorption of opposite chiral enantiomers between left and right-handed CPL[14,15], finds large employment in biosensing field because of its differential nature, free from background noise, and for the presence of several spectral features other than the typically broad plasmonic resonances[16–20].

Recently, ordered arrays of fully 3D chiral shapes have been realized and demonstrated optical resonances in the VIS spectral range[21,22]. In perspective of developing novel biosensing schemes, the helix geometry of such nanostructures, given the 3D nature, can have the additional advantage of increasing the available binding surface[23] exposed to the analyte, after suitable functionalization. Indeed, the performances of a sensing device are also intrinsically related to the functionalization protocol employed for specific biorecognition. Ideally, such a procedure should be chemically simple, cost-effective and not time-consuming, while ensuring stability, effective coverage and selectivity of the sensing nanostructures especially for analysis of complex biological matrix[24].

In this work, we demonstrate the potentiality of a 3D free-standing chiral metamaterial in a novel core-shell architecture, as a compact platform for optical biosensing detection reaching the femtomolar range. Here, the polymer-mediated surface functionalization is used as an effective approach for high-yield biorecognition, modulating, at the same time, the effective dielectric function of the sensing nanostructures.

The scheme of our biosensor is shown in figure 1. The engineered building block grown by Focused Ion Beam Induced Deposition (FIBID)[25] consists of a periodic array of chiral core-shell nanostructures, where the single element is a metallic helix. The chiral metamaterial is then prepared for biochemical functionalization through the conformal coverage with an ultrathin dielectric polymeric shell, the Poly-o-phenylediamine (P-oPD), followed by antibody covalent binding. The resulting core-shell architecture, with respect to dielectric function profile, represents the key element to achieve high sensitivity, because, on one side, it offers an ideal surface coverage for high stability molecular immobilization onto the nanohelices, while, on the other side, it allows to enhance the near- and far-field scattering of the nanohelices. This results in a large shift of the CD features to refractive index changes, and, consequently, to biodetection sensitivity in the femtomolar concentration range.

We applied our engineered sensing device against variable concentration of the transactive response (TAR) DNA-binding protein 43 (TDP-43), a distinctive protein of amyotrophic lateral sclerosis (ALS) and frontotemporal lobar degeneration (FTLD)[26,27]. Currently, the diagnosis of these neurodegenerative diseases is still clinically based and no biomarkers have yet been routinely incorporated into the clinical practice or clinical trials. TDP-43 is the main component of the pathological inclusions found in the cytoplasm of neurons and glial cells of the majority of ALS and Tau-negative FTLD cases. As a consequences TDP-43 has been largely proposed and studied as a potential biomarker for ALS and FTLD. Increased level of TDP-43 protein was found in cerebrospinal fluid (CSF) and plasma of patients with ALS and FTLD[28,29] using mass spectrometry[30] ,Western Blot and ELISA analysis[31,32]. However, the extremely low concentration of TDP-43 in biofluids represents a major limitation for its use as diagnostic biomarkers and requires the implementation of more challenging detection methods[29]. Recently, an ELISA test detected TDP-43 in CSF at concentration below 0.49 ng/mL[28] and a detection limit of 0.5ng/mL[33] has been achieved in serum using an electrochemical sensor. Our sensor demonstrated to be able to detect TDP-43 concentrations down to 10fM (corresponding to 0.43pg/mL). Moreover, according to specificity control experiments, our device scheme results robust against non-specific background noise in both dry and liquid environments, providing a fast and real-time detection even in complex body fluids like human serum.

The biosensor active area consists of a compact and periodic array of right-handed platinum nanohelices (Figure 2a). The geometrical and structural parameters have been engineered in our former works[21,34] to achieve chiro-optical effects in the visible spectral range. More details on fabrication process are reported in the Methods section.

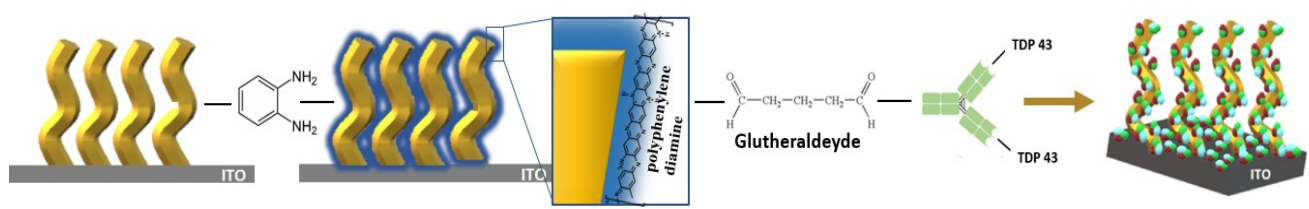

**Figure 1.** Scheme of the sensing device and functionalization protocol. The core-shell architecture arises from covering the fabricated helix array with the P-oPD insulating polymer. Then, the antibody is immobilized onto the shell after crosslinking by glutnereldeyde. Finally, the target analyte, deposited on the sample surface, is recognized by the specific antibody sites in dry environment.

Afterwards, we have exposed the nanohelices to oPD for polymerization by means of a cyclic voltammetry (CV) process (see methods) for the conformal deposition of a thin and compact dielectric polymer shell[35–37]. Such a functionalization strategy presents some structural advantages. First, the used monomer, the o-pheniledyamine (oPD), is characterized by a self-limiting polymeric growth under anodic oxidation in aqueous solution affording the formation of an ultrathin coating layer (few nanometers), where the aminic groups can be exploited to bind biomolecules. The self-limiting deposition is controlled by the polymer concentration, the buffer composition and the buffer pH in the CV process[38,39].

Second, the C affinity to the polymer ensures a homogeneous immobilization of molecules on the helix surface, consisting of an alloy of Pt crystalline nanograins uniformly embedded into an amorphous carbon matrix[21].

The shell morphology obtained on a Pt-nanohelix, grown on-purpose on a copper grid before and after P-oPD coating (see Methods for details on fabrication and CV), has been studied by STEM in Z-contrast mode, as shown in figure 2b, c,d. Figure 2b, obtained by combining both dark (DF) and bright field (BF) acquisition modes, evidences the conformal coverage of the low index polymeric shell around the helix core. High-magnification STEM images of the helix, acquired in dark field mode, before and after the deposition (figure 2c and 2d), show a very high uniformity of the polymeric thin shell throughout the whole helix structures beyond the shining helix core edges, (red arrows) with average thickness of 12±2 nm. The random dark/bright contrast visible into the helix core is related to the complex composition of the structure, with platinum grains appearing dark because of the high Z, and carbon matrix, with lower atomic number, appearing bright.

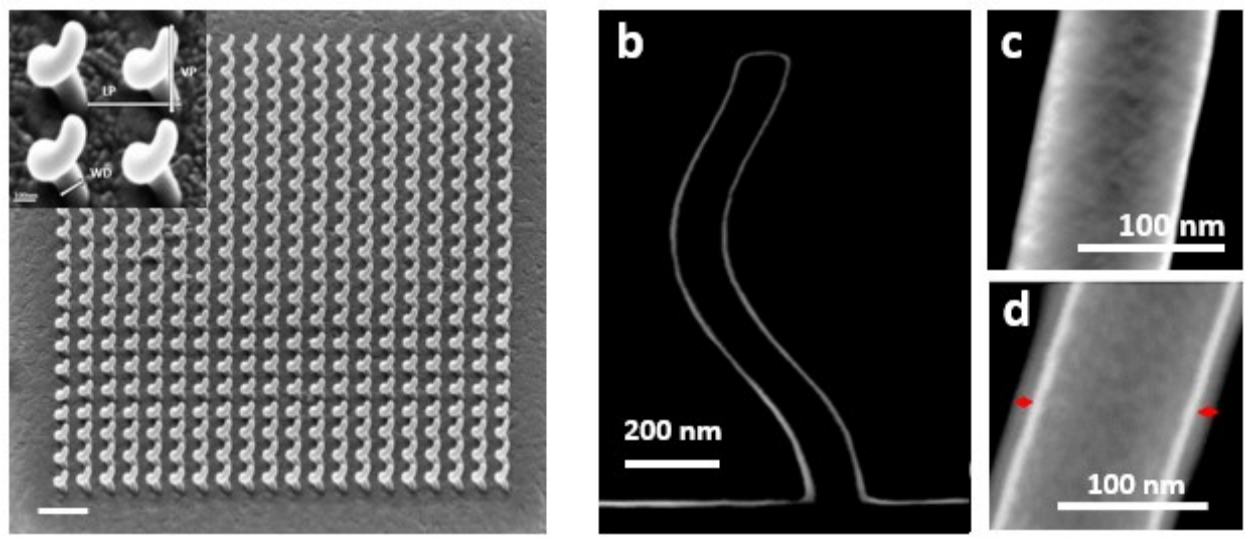

**Figure 2. a.** SEM image of nano-helix array based active area; the geometrical parameters showed in the inset are lateral period (LP) 500nm, vertical period (VP) ranging from 450 to 550nm, external diameter (ED) 310nm and wire diameter (WD) 120nm. The scale bar is 1um while the scale bar of the inset is 100nm. **b.** DF+BF High Angle Annular Magnification STEM of a core-shell Pt/P-oPD helix grown on a copper grid highlighting the thin and conformal nature of the P-oPD outer shell. **c, d.** STEM magnification of a nanohelix section before (c) and after (d) the P-oPD coating. The red arrows indicate the oPD shell thickness of (12±2) nm.

The circular dichroism spectra of both the core and the core-shell systems are calculated by the measured intensity of left and right-handed circularly polarized transmitted light ($T_{LCP}$ and $T_{RCP}$, respectively), according to:

$$CD = \left(\frac{\sqrt{T_{LCP}} - \sqrt{T_{RCP}}}{\sqrt{T_{LCP}} + \sqrt{T_{RCP}}}\right) \qquad (1)$$

The transmission measurements have been performed by means of a home-made confocal optical setup, described in the methods section.

The CD spectrum of the bare helix-based sensor exhibits two opposite dichroic bands (D1 and D2 of the blue line of figure 3a centered at $\lambda_M$=500nm and $\lambda_m$=840nm, respectively) due to hybridization phenomena occurring among the plasmonic resonances as a function of circularly polarized light along the chiral dipoles of the helix arms[40,41], with a zero dichroism point (ZDP) at $\lambda_{ZDP}$=600nm. After the shell deposition, a spectral redshift of the CD bands in the VIS occurs because the refractive index around the nanostructure increases (figure 3a red line). The unpatterned substrate remains barely affected by such a thin dielectric layer deposition (figure S1 in the supporting).

Core-shell nanosystems combining materials with different sign permittivities (i.e., metals with dielectrics) are expected to exhibit different effects on absorption and scattering at their resonance frequency, and to modify the electric field distribution around the nanostructures[42–44]. This depends on the relative core-shell size and on the interplay between nanostructure shape and size and inspecting wavelength. In our specific case, a decay length of 128nm is expected for the plasmonic field[45] at the CD peak of 500nm through the shell, by considering the optical dispersion of the Pt-based core[21] and of the P-oPD shell (supporting figure S2). Therefore, we can assume that the measured shell thickness of 12nm still supports the plasmon propagation out to the nanosystem surface.

Moreover, a better understanding of the polymer coating role on the optical response of the meta-sensor can be inferred by the analysis of the scattering signals and of the near field distribution for a single chiral nanohelix, before and after coating.

Figures 3 b-c show, respectively, the experimental scattering spectra of RCP and LCP light measured for the bare helix (blue line), and after the deposition of the outer shell (red line). Along with the slight spectral redshift, in line with what observed from the CD spectra, the core-shell architecture, when incident CPL matches the structure handedness, induces an enhancement of the scattering intensity, by a factor of 0.8 with respect to the bare core. This enhancement further increases by a factor of 2.7 with the opposite incident handedness. The results are in agreement with the numerical simulations of the scattering intensity in the supporting figure S3.

Furthermore, numerical analysis of the near field distribution for both, bare and core-shell nanosystems, at the resonance peaks for the two incident CPL components ($\lambda_{LCP}$ and $\lambda_{RCP}$), are reported in figure 3 d, e. In particular, the electric field hot spots close to the metal/air interface in the bare helix move towards the shell/air interface in the core-shell system. In such positions the field intensity is also increased. These two interrelated effects, the enhanced electric field and the increased far field scattering, could be attributed to the energy transfer between polarization charges of the dielectric shell and free electrons in the plasmonic core[46,47]. As seen later, with respect to biosensing application, the resulting interaction of polymer-coated helices with biomolecules can be more efficient, if compared with other monolayer-thin bioconjugation schemes.

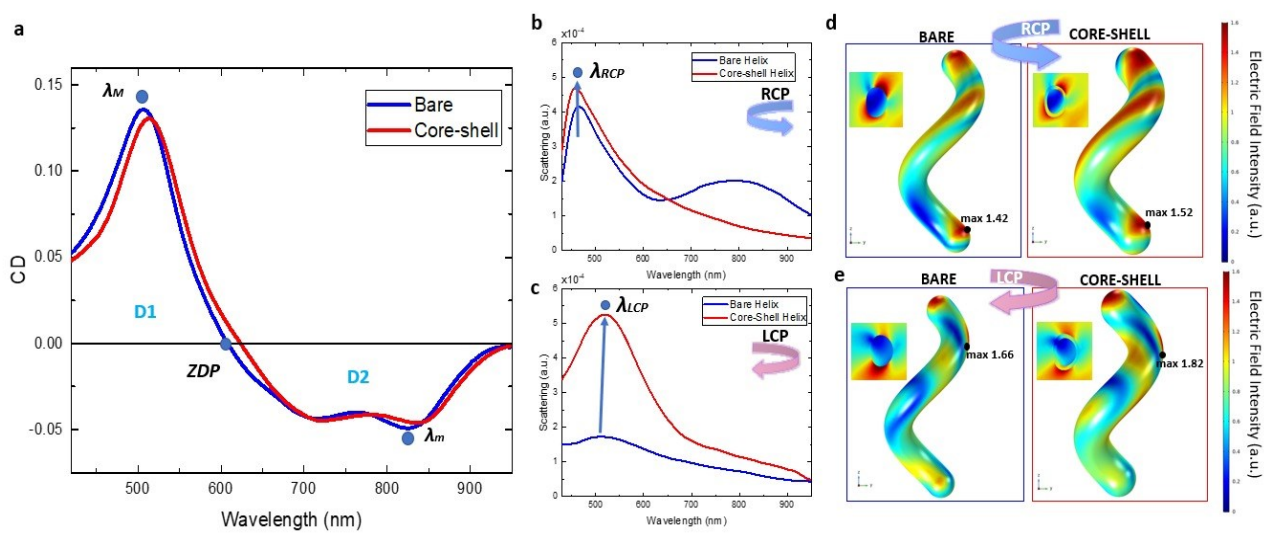

**Figure 3. a.** CD spectra of nano-helices array before and after shell coating. **b, c.** RCP (b) and LCP (c) scattering spectra measured for a single helix before and after the polymeric shell coating. The curves highlight a strong enhancement of the far field scattering for the core-shell architecture with respect to the core case, by a factor of 0.8 for RCP and 2.7 for LCP. **d, e.** Electric field distribution profile for both bare and core-shell single nano-helix calculated at the maximum resonance peak for RCP and LCP, respectively, showing a near field enhancement after shell coating. The insets show the cross section electric field intensity distribution in marked positions.

In order to evaluate the performance of our platform, first we monitored CD evolution of the core/shell system in liquid, by using glycerol–water mixtures with varying concentration from 0 to 20% (corresponding to a known refractive index range between 1.333 and 1.358 [48]).

Clearly, the RI increment of the external environment, from air (n=1) to water (n=1.333), induces a strong redshift of the CD spectrum toward NIR region (supporting figure S4a). Figures 4a, b, c show the shift of the main CD spectral signatures: $\lambda_M$, from 682nm to 690nm and ZDP, from 930nm to 949nm, respectively. The sensitivity, calculated from the linear fits as S=Δλ/Δn (where Δλ represents the wavelength shift and Δn the change of the refractive index of the glycerol-water solution), is reported for both $\lambda_M$ (squares) and ZDP (rhombs) in figure 4d. As expected[49,50], ZDP exhibits higher sensitivity (S=766nmRIU$^{-1}$) with respect to $\lambda_M$ (S=316nmRIU$^{-1}$). In addition, the smaller ZDP effective linewidth (δλ) allows to reach higher figure of merit (FOM =S/δλ) up to 1276 RIU$^{-1}$.

Here, we notice that, the analyte measurements at different concentration were performed in dry medium (air)[51,52]. Therefore, the RI calibration curves for $\lambda_M$ and for ZDP have been evaluated also considering this condition[53], giving from the linear fits similar values of slope (supporting figure S4). A comparison with state of art performances for LSPR-based systems[24] demonstrates the large potentiality for the proposed biosensing approach.

We completed the functionalization protocol by immobilizing the antibody with gluthereldeyde, acting as crosslinker agent between the aminic groups of the polymeric shell and the antibody. Then, the antigen TDP-43 has been incubated to allow the biorecognition event with the realized sensing layer (see methods for details).

Figure 5a shows the spectral redshift of CD curves, in linear correlation with molar concentration. We have measured the difference of the CD spectral features acquired after the antibody layer incubation and after the TDP-43 binding at different molar concentration, ranging from 1pM down to 10fM. ZDP (figure 5b) confirms a larger redshift as compared to the maximum $\lambda_M$ and the minimum $\lambda_m$ CD features (figure S5). The linear trend of the spectral shifts as a function of the molar concentration is noticeable in figure 5c. The covered range fulfils the early diagnosis requirements of this family of biomarkers[29,33].

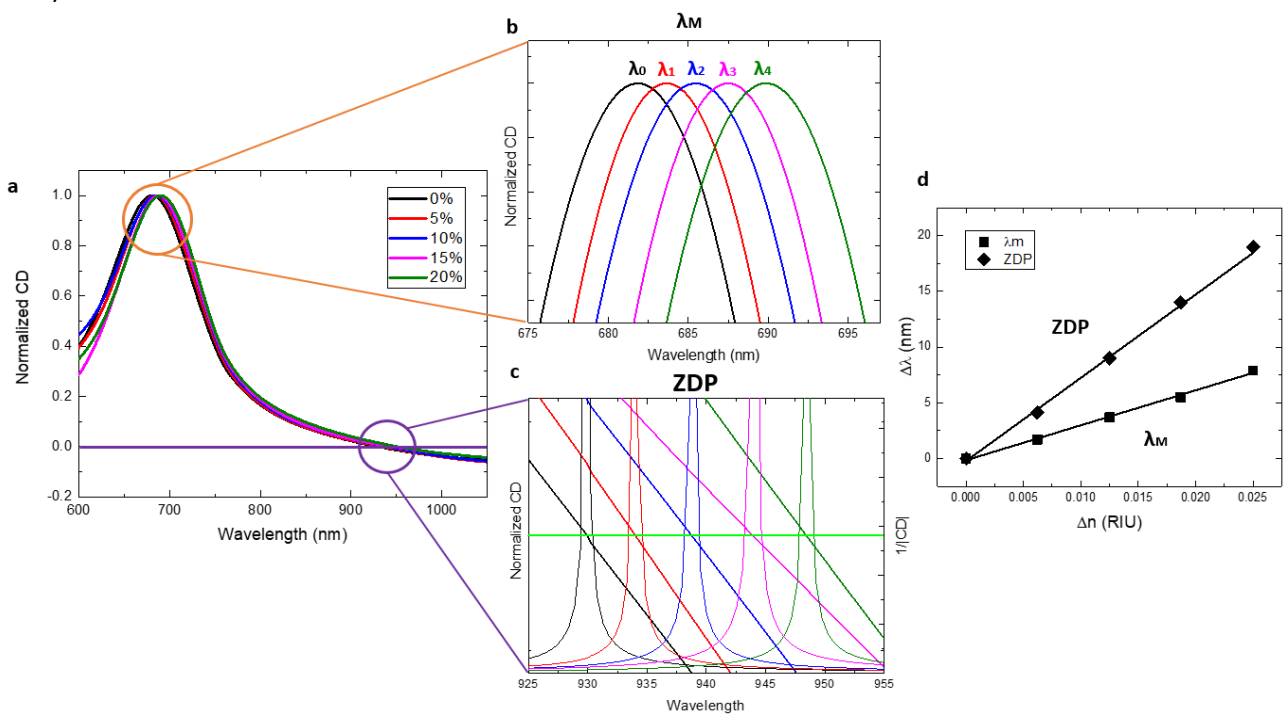

**Figure 4. a.** Normalized CD spectra of the core-shell nano-helices immersed in a glycerol-water solution at different molar concentrations. **b, c.** High-magnification CD spectra for $\lambda_M$ (indicated with the orange circle) and ZDP (indicated with purple circle). In particular, 1/|CD| is calculated for ZDP in order to evaluate the FWHM values. **d.** Relationship between $\lambda_M$ (black square symbols) and ZDP (black rhombus symbols), and the refractive index. The standard deviation (below 0.4nm) retrieved for the data points falls within the size of the symbols.

The real effectiveness of our functionalization method has been tested by comparing the TDP-43 detection results with the standard functionalization method based on the self-assembly monolayer of thiols, commonly used for metallic surfaces including platinum[54,55]. In the thiols- based functionalization experiment, the detected ZDP redshift (figure S6) was 5nm, five-times lower than the value obtained with the core-shell architecture (27nm) for the same analyte concentration of 1pM. As anticipated before, this results from the worthy near- and far- field scattering intensity obtained with the conformal polymer. Moreover, unlike thiols layer, that binds only on the Pt grains in the complex helix material alloy, while leaving the carbon surface fraction uncovered, the polymeric shell conformally coats the entire helix surface. This allows to maximize the total specific binding sites available for the target analyte and amplify the detected signal.

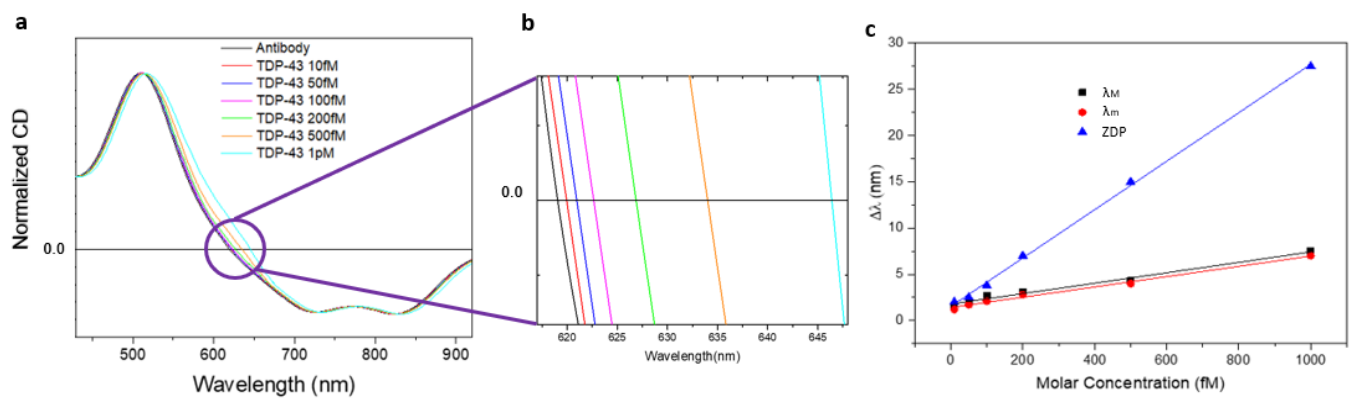

**Figure 5. a**. Normalized CD spectra acquired after the only deposition of the antibody layer (spectral reference) and after different concentrations of TDP-43; **b.** the high-magnification around the ZDP region allows to evaluate the spectral shift of the crossing point between antibody and the antigen at different molar concentrations; **c.** Linear fit of the spectral shift for $\lambda_M$, $\lambda_m$ and ZDP respectively. The size of data points represents the error bars.

The sensor specificity has been tested performing two control experiments. In the first case, we have used our nanohelix-based sensor with immobilized antibodies for TDP-43 on the polymeric shell, to reveal the non-specific detection of Tau protein, a biomarker related to AD and PD neurodegenerative diseases [56]. The sensor has been incubated with a solution of Tau protein at 500 fM molar concentration. As shown in figure 6a that compares the chiroptical measurements before and after the antigen deposition, the multiple CD spectral features show no distinct variation on addition of non-target Tau protein, because no specific binding events occur, demonstrating that the nano sensor is selective and specific to target molecules.

The second experiment consists to detect the TDP-43 in the complex environment of human serum, characterized by the presence of different biomolecules that can potentially interfere with the specific analyte detection. The results are shown in figure 6b, where the CD features are recorded (i) with the only immobilized TDP-43 antibody, (ii) after the addition of human serum and (iii) after the addition of human serum spiked with TDP-43 (at 500fM), respectively. While the human serum acting as background does not induce distortion in the CD signatures, the addition of the spiked serum leads to significant ZDP redshift of about 15nm, the same value recorded in the detection of pure TDP-43 with the same concentration (500 fM) (figure 5).

Moreover, the possibility offered by our chiral sensor to use the CD spectrum introduces the additional property of the unperturbability to signals coming outside the active area. Indeed, considering that the thin layer of oPD is also deposited on the substrate, it creates binding sites for antibody-antigen pairs, generating the same transmission offset for both the circularly polarized transmitted lights, that the CD can delete because it is a differential signal. These results point out the excellent and stable performances and the reliability of the device.

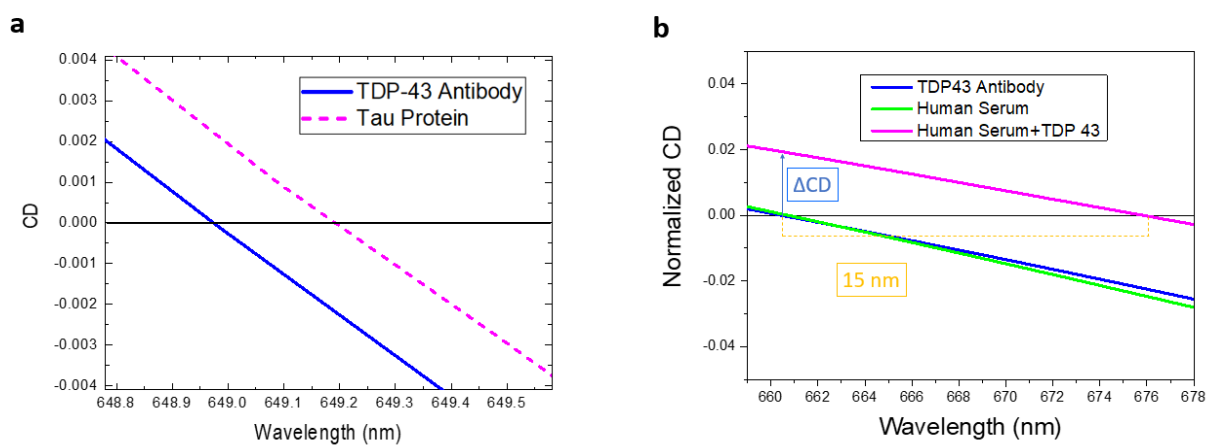

**Figure 6. a.** Detailed plots of the ZDP for CD spectra acquired after the incubation of TDP-43 antibody (blue curves) and Tau protein (pink dashed line). As depicted by ZDP position, no significant changes are recorded **b.** Detailed plots of the ZDP for CD spectra measured after the incubation of TDP-43 antibody (blue curves), human serum only (green line) and human serum + TDP 43 (pink lines). Large spectral shift of 15nm is observed only in presence of the target antigen.

Our work proposes a compact metamaterial with chiral properties as a novel optical biosensing concept, with sensitivity of about 800nm/RIU and FOM= 1276 RIU$^{-1}$, and capable to detect TDP-43 protein concentrations between 1pM and 10fM, even in the complex human serum fluid, thanks to the background-free differential signal. Further improvement in sensitivity can be envisioned thanks to the size-, geometry- and material-dependent engineering of the sensor [45,57], leading to even more pronounced bisignated circular dichroism (in the degree level and beyond), with steeper stop bands, for larger and easier-to-detect spectral shifts. Being a miniaturized photonic component, it can be embedded within a portable point of care system, with a simple integrated optical read out, considering that it works in transmittance under normal incidence conditions, thus not requiring complex excitation geometries. Moreover, the intrinsic chirality of the nanohelices can open perspectives in enantiomeric detection, of critical importance for chemical and pharmaceutical applications. Furthermore, the finding that the core-shell architecture, used for biochemical functionalization, can enhance the near- and far-field scattering properties opens wide perspectives to explore novel nanophotonics schemes, not only for sensing but also for fundamental light-matter interactions.

**Experimental Section**

*Sample Fabrication:* A series of arrays of 3D Platinum-based Nano-helices single loop with lateral period (LP) 500nm, vertical period (VP) ranging from 450 to 550nm, external diameter (ED) 300nm and wire radius (WR) 90-120nm, were realized on ITO-on-glass substrate with Focused Ion Beam Induced Deposition technique employing a Carl Zeiss Auriga40 Crossbeam FIB/SEM system. This system allows the realization of 3D structures together with a gas injection system that contains the source of trimethyl-methylcyclopentadienyl-platinum(IV) precursor. The gas is injected in the chamber and the ion beam (with parameters 1pA, 30KeV acceleration voltage and 10nm step size) dissociates the gas molecules to obtain, locally, a controlled and uniform growth of the nano helices array, sized 10x10µm. The vacuum chamber was kept within a pressure range from $8.80 \times 10^{-7}$ mbar to $9 \times 10^{-6}$ mbar during the deposition time. To perform the STEM characterization a single helix was grown on-purpose on a copper grid with the same growth conditions of the array.

*Optical characterization*: Transmission spectra were recorded with an optical microscope Zeiss Axioscope A1 with a home-made confocal system coupled to an imaging spectrometer. Light from a tungsten lamp is focused on the sample with a condenser with NA<0.1 and it is collected using a 40x objective lens with NA<0.95. Then, the light transmitted through the sample is guided through a system made by three lenses that reconstruct, collimate and refocus the real space image. The selected real image is reconstructed and directed to a CCD camera (Hamamatsu Orca R2) coupled with a 200 mm spectrometer for measurements in the visible spectral range and to an InGaAs detector (Princeton Instruments, OMA V InGaAs linear array) coupled with a 300 mm spectrometer in the near-infrared region of the spectrum. Adjustable square slits were used to select the array area. The circularly polarized light has been produced using a combination of a linear polarizer and a quarter-wave plate. For measurements in visible spectral range a linear polarizer and a superachromatic waveplate (Carl Zeiss, 400-800nm) have been used while for measurements in NIR region a linear polarizer (Thorlabs, LPVis100-MP 550nm - 1.5µm) coupled to an achromatic quarter wave plate (Newport, Achromatic Waveplate 700-1000nm) have been used. All transmission measurements were normalized to the optical response of the substrate. The scattering spectra of single chiral nanostructure have been performed with a confocal spectrometer (HORIBA LabRam HR) equipped with a cooled charge-coupled device (CCD) Camera (Andor DU 420A-OE 325) and 150 g/mm diffraction grating. The position and visualization of the sample are carried out by another CCD camera (Cannon 400D). For the measurements the following darkfield scheme is implemented. The non-polarized light from a halogen lamp (Ocean Optics HL-2000-HP) connected with 400 µm Vis-NIR fiber (Ocean Optics), passes through a linear polarizer and superachromatic quarter-wave plate (Thorlabs, 325-1100 nm). Then the incident light is focused by an objective (Mitutoyo Plan Apo NIR, infinity corrected, 10×, NA = 0.26) and obliquely excite the sample with helices at the angle of 65 degrees to the surface normal. The scattered signal is collected by the second achromatic objective (50×, NA = 0.42 Mitutoyo Plan Apo NIR) placed perpendicularly to the sample surface. All scattering spectra are normalized to the lamp spectra measured at the same experimental condition.

*Polymer Electrodeposition*: The polymer is deposited on the helices surface from a solution of o-PD (SIGMA Aldrich) at a concentration of 0,01mg/mL in the acetate buffer (pH 5) by cyclic voltammetry (with potential ranging from -0.02V to 0.8V, scan rate 50mV/s) with an Autolab PGSTAT 302N. The ITO samples with helix arrays were cut in size 1x1 cm before immersion in the solution.

*SEM and STEM images*: SEM and STEM characterization was performed by means of a Merlin Zeiss microscope operating in scanning mode on helix array and single helix fabricated on a TEM grid copper by combining dark-field and bright field. In order to obtain Z contrast sensitiveness from the images, STEM was configured in high-angle annular dark-field mode.

*Numerical simulation:* Finite element method (FEM) simulation was developed through Comsol Multiphysics 5.4, by exploiting wave optics module and carrying out a frequency domain study of electromagnetic waves. All the physical dimensions were evaluated by realizing a geometry consisting of the Pt/C helix, the oPD shell, the air medium around the structure and a perfectly matched layer. Wavelength sweeping was performed in a range between 400 nm and 1000 nm with a step of 10nm, by exciting the structures with circularly polarized light through a background field directed with zenith distance of p/3, then collecting far field signal along zenith, in a solid angle of 0.7\π steradians, according to the experimental setup. The material dispersions used are from [21] for platinum/carbon alloy and from Supporting Fig. S2 for the polymer shell.

*Functionalization procedure:* To enable the biorecognition assay, the developed functionalization protocol, after the polymer coating, foresees the incubation with a solution of glutaraldehyde (SIGMA Aldrich) 0,1% (w/v) for 40 minutes: the glutaraldehyde acts as a crosslinker between the P-oPD and antibody's amino groups. Later, the immobilization of polyclonal antibody directed against the C-terminal amino acids of human TDP-43 (Proteintech) was carried out through 1h incubation, with an antibody solution of 2ng/ml in humid conditions in order to avoid the evaporation of the sample drop. Then, the platform was rinsed with water and dried under nitrogen flow. In the control experiments with thiols functionalization the antibodies were immobilized to a self assembled monolayer realized by the deposition of a mercaptoundecanoic acid (11-MUA) (SIGMA Aldrich) (0.2 mM) in ethanol incubating the sensors with the thiols'solution for 2h. Then the COOH groups of MUA were activated by incubation with a solution of N-hydroxysuccinimide (NHS) (SIGMA Aldrich) and N-ethyl- N-(3-dimethylaminopropyl carbodiimide hydrochloride (EDC) (SIGMA Aldrich) at a ratio of 1:4 in milliQ water for 30 min. Then a solution of antibodies in PBS at a final concentration of 2ng/ml was incubated for 1 h to allow a covalent binding to SAM. To allow the biorecognition between the antibody and the analyte, the functionalized sensor were incubated with solutions of the un-tagged human recombinant TDP-43 protein (amino acids 1-414; Novus Biological) at different concentrations ranging from 1pM (43pg/mL) to 10 fM (0.43pg/mL) in PBS for 30 minutes, then rinsed in milliQ water and finally dried. The incubation process and optical measures were repeated for increasing concentrations of TDP-43. For the specificity tests the functionalized sensor has been incubated with solutions of Tau protein (SIGMA Aldrich) at 500fM in PBS for 30 minutes, then rinsed in milliQ water and finally dried. For the complex environment experiment, the functionalized sensor has been incubated with solutions of TDP-43 (Novus Biological) at 500fM in human serum for 30 minutes, then rinsed in PBS and finally dried. Serum blood sample had been obtained after signature of the informed consent, and stored in the Institute of Experimental Neurology (INSPE) biobank. All experimental protocol was approved by San Raffaele Scientific Institute Ethical Committee (EC: 106/INT/2018) (Milan, Italy), and has, therefore, been performed in accordance with the ethical standards laid down in the 1964 Declaration of Helsinki and its later amendments. After optical measurements, the sensors can be regenerated through UV-ozone exposure followed by ethanol rinsing, leading to a weakening and complete remotion of the polymer chain with the linked antibody-antigen pair and returning to original CD spectral features.

**Acknowledgements**

This work was supported by ''Tecnopolo per la medicina di precisione'' (TecnoMed Puglia) – Regione Puglia: DGR no. 2117 del 21/11/2018 CUP: B84I18000540002. V.T. thanks the Short Term Mobility program of CNR. Iolena Tarantini for technical support.

**References**

[1]   A. G. Brolo, *Nat. Photonics* **2012**, *6*, 709.

[2]   A. P. F. Turner, *Chem. Soc. Rev.* **2013**, *42*, 3184.

[3]   J. R. Mejía-Salazar, O. N. Oliveira, *Chem. Rev.* **2018**, *118*, 10617.

[4]     J. N. ANKER, W. P. HALL, O. LYANDRES, N. C. SHAH, J. ZHAO, R. P. VAN DUYNE, in *Nanosci. Technol.*, Co-Published With Macmillan Publishers Ltd, UK, **2009**, pp. 308–319.

[5]     M. I. Stockman, *Science (80-. ).* **2015**, *348*, 287 LP.

[6]     J. Liu, M. Jalali, S. Mahshid, S. Wachsmann-Hogiu, *Analyst* **2020**, *145*, 364.

[7]     C.-Y. Chang, H.-T. Lin, M.-S. Lai, T.-Y. Shieh, C.-C. Peng, M.-H. Shih, Y.-C. Tung, *Sci. Rep.* **2018**, *8*, 11812.

[8]     K. V. Sreekanth, Y. Alapan, M. Elkabbash, E. Ilker, M. Hinczewski, U. A. Gurkan, A. De Luca, G. Strangi, *Nat. Mater.* **2016**, *15*, 621.

[9]     K. S. Lee, M. A. El-Sayed, *J. Phys. Chem. B* **2006**, *110*, 19220.

[10]    D. E. Charles, D. Aherne, M. Gara, D. M. Ledwith, Y. K. Gun'ko, J. M. Kelly, W. J. Blau, M. E. Brennan-Fournet, *ACS Nano* **2010**, *4*, 55.

[11]    M. R. Gartia, A. Hsiao, A. Pokhriyal, S. Seo, G. Kulsharova, B. T. Cunningham, T. C. Bond, G. L. Liu, *Adv. Opt. Mater.* **2013**, *1*, 68.

[12]    Z. Wang, F. Cheng, T. Winsor, Y. Liu, *Nanotechnology* **2016**, *27*, 1.

[13]    J. T. Collins, C. Kuppe, D. C. Hooper, C. Sibilia, M. Centini, V. K. Valev, *Adv. Opt. Mater.* **2017**, *5*, 1.

[14]    X. Wang, Z. Tang, *Small* **2017**, *13*, 1.

[15]    M. Schäferling, D. Dregely, M. Hentschel, H. Giessen, *Phys. Rev. X* **2012**, *2*, 1.

[16]    S. Yoo, Q. H. Park, *Nanophotonics* **2018**, 1.

[17]    E. Hendry, T. Carpy, J. Johnston, M. Popland, R. V. Mikhaylovskiy, A. J. Lapthorn, S. M. Kelly, L. D. Barron, N. Gadegaard, M. Kadodwala, *Nat. Nanotechnol.* **2010**, *5*, 783.

[18]    J. Garciá-Guirado, M. Svedendahl, J. Puigdollers, R. Quidant, *Nano Lett.* **2018**, *18*, 6279.

[19]    N. Claes, A. L. Cortajarena, E. López-Martínez, H. Eraña, S. Bals, J. Kumar, J. Castilla, V. F. Martín, D. M. Solís, L. M. Liz-Marzán, *Proc. Natl. Acad. Sci.* **2018**, *115*, 3225.

[20]    M. Hentschel, M. Schäferling, X. Duan, H. Giessen, N. Liu, *Sci. Adv.* **2017**, *3*, 1.

[21]    M. Esposito, V. Tasco, F. Todisco, M. Cuscunà, A. Benedetti, M. Scuderi, G. Nicotra, A. Passaseo, *Nano Lett.* **2016**, *16*, 5823.

[22]    J. G. Gibbs, A. G. Mark, S. Eslami, P. Fischer, *Appl. Phys. Lett.* **2013**, *103*, 213101.

[23]    G. Palermo, G. E. Lio, M. Esposito, L. Ricciardi, M. Manoccio, V. Tasco, A. Passaseo, A. De Luca, G. Strangi, *ACS Appl. Mater. Interfaces* **2020**, *12*, 30181.

[24]    T. Xu, Z. Geng, *Biosens. Bioelectron.* **2021**, *174*, 112850.

[25]    M. Manoccio, M. Esposito, A. Passaseo, M. Cuscunà, V. Tasco, *Micromachines* **2021**, *12*, 1.

[26]    R. L. French, Z. R. Grese, H. Aligireddy, D. D. Dhavale, A. N. Reeb, N. Kedia, P. T. Kotzbauer, J. Bieschke, Y. M. Ayala, *J. Biol. Chem.* **2019**, *294*, 6696.

[27]    F. Hans, M. Eckert, F. von Zweydorf, C. J. Gloeckner, P. J. Kahle, *J. Biol. Chem.* **2018**, *293*, 16083.

[28]    T. Kasai, T. Tokuda, N. Ishigami, H. Sasayama, P. Foulds, D. J. Mitchell, D. M. A. Mann, D. Allsop, M. Nakagawa, *Acta Neuropathol.* **2009**, *117*, 55.

[29]    E. Feneberg, E. Gray, O. Ansorge, K. Talbot, M. R. Turner, *Mol. Neurobiol.* **2018**, *55*, 7789.

[30]    F. Kametani, T. Obi, T. Shishido, H. Akatsu, S. Murayama, Y. Saito, M. Yoshida, M. Hasegawa, *Sci. Rep.* **2016**, *6*, 23281.

[31]    P. Foulds, E. McAuley, L. Gibbons, Y. Davidson, S. M. Pickering-Brown, D. Neary, J. S. Snowden, D. Allsop, D. M. A. Mann, *Acta Neuropathol.* **2008**, *116*, 141.

[32]  E. Verstraete, H. B. Kuiperij, M. M. van Blitterswijk, J. H. Veldink, H. J. Schelhaas, L. H. van den Berg, M. M. Verbeek, *Amyotroph. lateral Scler. Off. Publ. World Fed. Neurol. Res. Gr. Mot. Neuron Dis.* **2012**, *13*, 446.

[33]  Y. Dai, C. Wang, L.-Y. Chiu, K. Abbasi, B. S. Tolbert, G. Sauvé, Y. Yen, C.-C. Liu, *Biosens. Bioelectron.* **2018**, *117*, 60.

[34]  M. Esposito, V. Tasco, F. Todisco, A. Benedetti, I. Tarantini, M. Cuscunà, L. Dominici, M. De Giorgi, A. Passaseo, *Nanoscale* **2015**, *7*, 18081.

[35]  J. W. Long, C. P. Rhodes, A. L. Young, D. R. Rolison, *Nano Lett.* **2003**, *3*, 1155.

[36]  U. Olgun, M. Gülfen, *React. Funct. Polym.* **2014**, *77*, 23.

[37]  I. Losito, F. Palmisano, P. G. Zambonin, *Anal. Chem.* **2003**, *75*, 4988.

[38]  R. Mazeikiene, A. Malinauskas, *Synth. Met.* **2002**, *128*, 121.

[39]  I. Losito, E. De Giglio, N. Cioffi, C. Malitesta, *J. Mater. Chem.* **2001**, *11*, 1812.

[40]  Y. R. Li, R. M. Ho, Y. C. Hung, *IEEE Photonics J.* **2013**, *5*, 2700510.

[41]  Z.-Y. Zhang, Y.-P. Zhao, *Appl. Phys. Lett.* **2007**, *90*, 221501.

[42]  A. Abrashuly, C. Valagiannopoulos, *Phys. Rev. Appl.* **2019**, *11*, 14051.

[43]  A. Sheverdin, C. Valagiannopoulos, *Phys. Rev. B* **2019**, *99*, 75305.

[44]  S. A. Hassani Gangaraj, C. Valagiannopoulos, F. Monticone, *Phys. Rev. Res.* **2020**, *2*, 23180.

[45]  M. Cuscunà, M. Manoccio, M. Esposito, M. Scuderi, G. Nicotra, I. Tarantini, A. Melcarne, V. Tasco, M. Losurdo, A. Passaseo, *Mater. Horizons* **2021**, *8*, 187.

[46]  P. Yu, Y. Yao, J. Wu, X. Niu, A. L. Rogach, Z. Wang, *Sci. Rep.* **2017**, *7*, 7696.

[47]  D. Avşar, H. Ertürk, M. P. Mengüç, *Mater. Res. Express* **2019**, *6*, 65006.

[48]  N.-S. Cheng, *Ind. Eng. Chem. Res.* **2008**, *47*, 3285.

[49]  H. H. Jeong, A. G. Mark, M. Alarcón-Correa, I. Kim, P. Oswald, T. C. Lee, P. Fischer, *Nat. Commun.* **2016**, *7*, DOI 10.1038/ncomms11331.

[50]  T. V. A. G. de Oliveira, K. E. Gregorczyk, M. Knez, P. Vavassori, J. Åkerman, M. Kataja, Z. Pirzadeh, S. van Dijken, N. Maccaferri, A. Dmitriev, *Nat. Commun.* **2015**, *6*, 1.

[51]  F. Yesilkoy, R. A. Terborg, J. Pello, A. A. Belushkin, Y. Jahani, V. Pruneri, H. Altug, *Light Sci. Appl.* **2018**, *7*, 17152.

[52]  A. L. Hernández, R. Casquel, M. Holgado, I. Cornago, F. Fernández, P. Ciaurriz, F. J. Sanza, B. Santamaría, M. V Maigler, S. Quintero, M. F. Laguna, *Sensors Actuators B Chem.* **2018**, *259*, 956.

[53]  Y. Shen, J. Zhou, T. Liu, Y. Tao, R. Jiang, M. Liu, G. Xiao, J. Zhu, Z.-K. Zhou, X. Wang, C. Jin, J. Wang, *Nat. Commun.* **2013**, *4*, 2381.

[54]  S. E. Eklund, D. E. Cliffel, *Langmuir* **2004**, *20*, 6012.

[55]  B. V Petrovic, M. I. Djuran, Z. D. Bugarcic, *Met. Based. Drugs* **1999**, *6*, 521864.

[56]  T. F. Gendron, L. Petrucelli, *Mol. Neurodegener.* **2009**, *4*, 13.

[57]  M. Esposito, V. Tasco, F. Todisco, M. Cuscunà, A. Benedetti, D. Sanvitto, A. Passaseo, *Nat. Commun.* **2015**, *6*, 1.

**Supporting information**

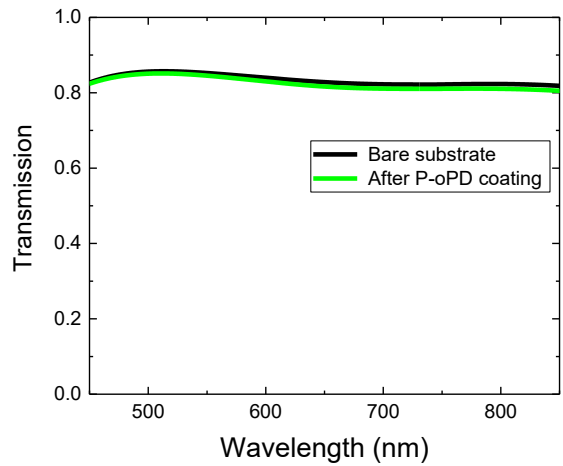

Figure S1. Transmission spectra of the ITO substrate before and after the shell coating.

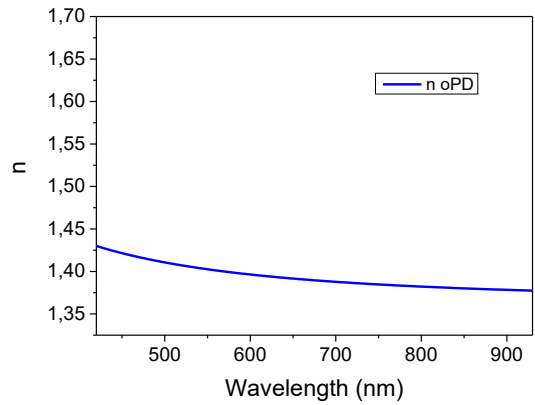

Figure S2. Retrieved refractive index of P-oPD deposited at the same conditions used for depositing the polymeric shell around the helix on an ITO substrate.

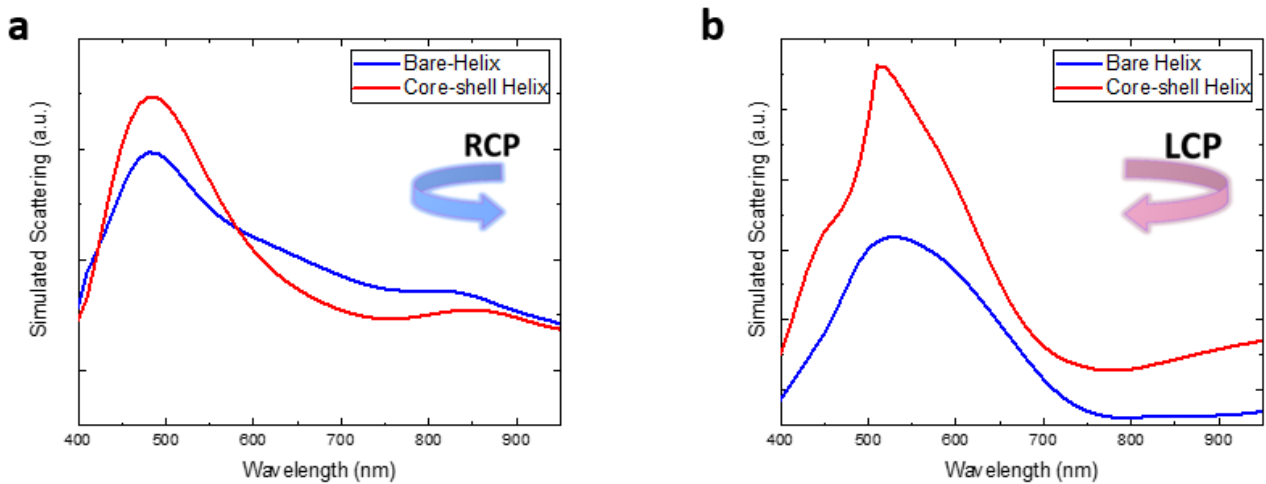

Figure S3. Simulated scattering spectra for a single bare helix (blue line) and core-shell helix (red line) for both circular polarization handedness: RCP (a) and LCP (b), respectively.

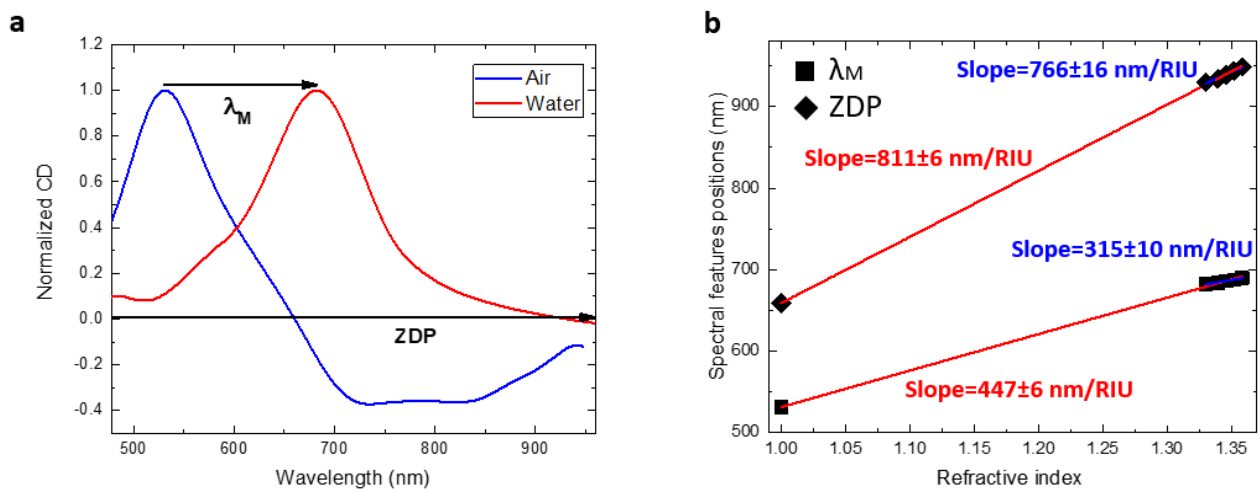

Figure S4. a) CD spectra of the measured core-shell nano-helices arrays in air (blue line) and water (red line). The large redshift is related to the RI of the different environment: air n=1 and water n=1.334. b) Trend of the spectral position of λM (Black square) and ZDP (black rhombs) measured in air and for glycerol-water solutions at different concentration corresponding to different RI variations from n=1.333 (100% water, 0% glycerol) to n=1.346 (80% water, 20% glycerol). The lines correspond to their respective calculated linear fitting. Particularly the linear fitting of the points calculated in the refractive index range from n=1.333 to n=1.346 (blue line) gives rise to an index sensitivity of 315 nm/RIU for λM and 766 nm/RIU for ZDP. The sensitivity obtained from the linear fitting in the RI range from n=1 to n=1.346 (red line) is instead 447nm/RIU and 811 nm/RIU. The slope value retrieved especially in the case of ZDP is comparable (and even slightly improved) with and without considering their spectral position measured in air environment and the entire spectral range can be well fitted linearly [1].

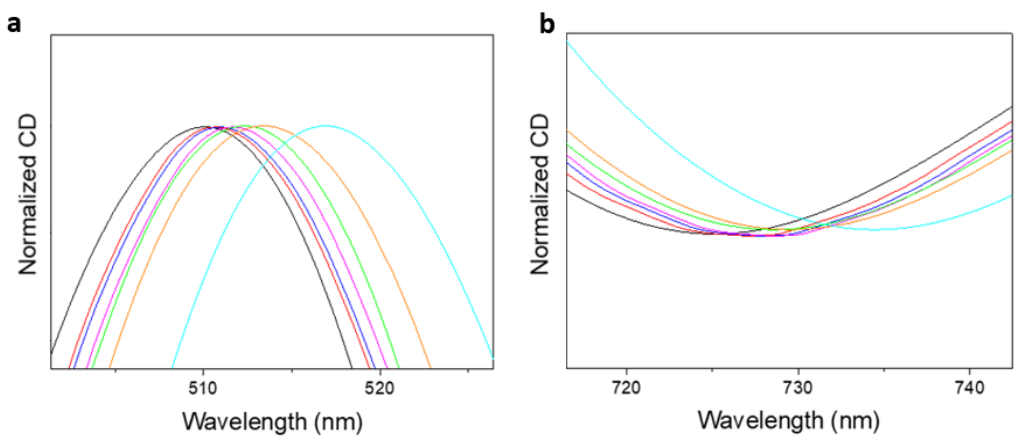

Figure S5. Detailed plots of the resonance shifts of the CD at λM (a) and λm (b), respectively.

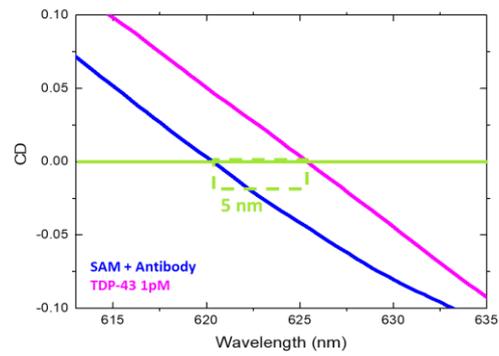

Figure S6. Shift of the crossing point between antibody and the antigen acquired by performing a previous functionalization with SAM technique. As one can observe, the redshift is less than 5 times smaller than the system we developed.

[1] Y. Shen, J. Zhou, T. Liu, Y. Tao, R. Jiang, M. Liu, G. Xiao, J. Zhu, Z.-K. Zhou, X. Wang, C. Jin, J. Wang, *Nat. Commun.* **2013**, *4*, 2381.